# How reliable are the Powell-Wetherall plot method and the maximum-length approach? Implications for length-based studies of growth and mortality

**Ralf Schwamborn**

Oceanography Dept., Federal University of Pernambuco (UFPE), 50670-901 Recife, Brazil. e-mail: rs@ufpe.br

## Abstract

Length-based methods are the cornerstone of many population studies and stock assessments. This study tested two widely used methods: the Powell-Wetherall (P-W) plot and the $L_{max}$ approach, i.e., estimating $L_\infty$ directly from $L_{max}$. In most simulations, P-W estimates of the ratio total mortality / growth (Z/K ratio) were biased beyond acceptable limits (bias > 30%). Bias in Z/K showed a complex behavior, without possible corrections. Estimates of asymptotic length ($L_\infty$) were less biased than Z/K, but were very sensible to intra-cohort variability in growth and to changes in the occurrence of large individuals in the sample. Exclusion of the largest size classes during the regression procedure or weighing by abundance does not solve these issues. Perfect linearization of the data and absurdly narrow confidence intervals for Z/K will lead users to erroneous overconfidence in outputs. Clearly, the P-W method is not suitable for the assessment of Z/K ratios of natural populations. Estimation of $L_\infty$ may be tentatively possible under very specific conditions, with necessary external verifications. Also, this study demonstrates that there is no way to estimate $L_\infty$ directly from $L_{max}$, since there is no particular relationship to expect *a priori* between $L_\infty$ and $L_{max}$. Errors in estimating $L_\infty$ directly affect the estimate of the growth constant K and all other subsequent calculations in population studies, stock assessments and ecosystem models. New approaches are urgently needed for length-based studies of body growth (e.g., unconstrained curve fit with subsequent bootstrapping), that consider the inherent uncertainty regarding the underlying data and processes.

# Introduction

*Length-based methods in fisheries science*

Length-based methods have been used to infer body growth in aquatic animals since the very beginning of fisheries science (Petersen, 1891). One particularly popular length–based method is the Powell-Wetherall (P-W) approach (Wetherall, 1986). It is a central part of the FISAT II package (Gayanilo et al., 1995, 2005, available at www.fao.org/fi/statist/fisoft/fisat), a widely used software package for stock assessment and population dynamics for fish and invertebrates worldwide. The P-W plot is the most commonly used method for length-based analyses worldwide, especially for tropical fish and invertebrate stocks (e.g., Arellano, 1989, Anbalagan et al., 2016). It has been used to infer growth, mortality and vulnerability of hundreds of economically relevant fish stocks (e.g., Lucena-Frédou et al., 2017), and numerous invertebrate species, including such data-poor organisms as squids (Mohamed and Rao, 1997), land crabs (Silva et al., 2014), and even jellyfish (Palomares and Pauly, 2009a).

Virtually all length-based studies performed in the last two decades have been based on FISAT II (Gayanilo et al., 1995, 2005) and its fast, simple, and straightforward length-based methods. Recently, these methods have been incorporated into R, within two new packages called "ELEFAN in R" (Pauly and Greenberg, 2013) and "TropFishR" (Mildenberger et al., 2017). These new software packages contain many promising new features, but still include the P-W method as a central part of the proposed analyses (Mildenberger, 2017).

*Fitting the growth function*

All length-based methods start with the fit of a growth function. Usually, the von Bertalanffy Growth Function (VBGF) is used for this purpose (von Bertalanffy, 1934, 1938). The shape of the VBGF is mainly determined by two parameters: the growth constant K (that determines its curvature, high K meaning fast growth) and the asymptotic length $L_\infty$, (that is, the mean length of a theoretical cohort with infinite age and zero mortality, a higher $L_\infty$ together with low mortality generally means larger fish at maximum age). The third parameter of this function is $t_0$, which defines the horizontal position of the curve.

A simple approach is generally recommended to fit the VBGF to length data (Gayanilo et al., 1995, 2005, Mildenberger, 2017). The first step is to assess $L_\infty$, either by using the Powell-Wetherall (P-W) method (Wetherall, 1986, Pauly, 1986), or simply based on the largest organism in the sample ("$L_{max}$ approach" or maximum-length approach, Mathews and Samuel, 1990). Wetherall et al. (1987) saw the simplicity and visually intuitive shape of their regression-based method as a big advantage over non-regressive length-based methods such as the Beverton and Holt (1956) and Powell (1979) methods.

The perceptions that $L_\infty$ can be easily estimated and that it is always close to $L_{max}$ (i.e., the length of largest individual in a given dataset) are steadfast paradigms of fisheries science. The recommendation to estimate $L_\infty$ directly from $L_{max}$ is part of many classic papers and standard textbooks and tutorials, e.g., "the largest fish or the average of the ten largest fish can be used for a small or large sample, respectively" (Wetherall et al., 1987) or "… you may simply use the largest fish." (Sparre and Venema, 1998).

The determination of $L_\infty$ is of paramount importance for the assessment of body growth. For instance, the estimate for K, when obtained from LFDs (length-frequency distributions), is determined by the initial value of $L_\infty$. A larger $L_\infty$ will automatically lead to a lower estimate of K, and vice versa. The first step (i.e., the estimation of $L_\infty$) determines all further steps and parameters when inferring growth models from length data and directly affects all subsequent population models.

*The Powell-Wetherall plot*

The P-W method is based on a linearizing transformation of size classes. There are two versions of it. The version of the plot used in FISAT II and TropFishR is the "modified" version, i.e., a version where the method originally proposed by Wetherall (1986) was modified by Pauly (1986) to allow for a more simple estimation of $L_\infty$. Pauly and Greenberg (2013) later returned to the original, "unmodified" version of the P-W plot for the software package "ELEFAN in R".

The basic idea of the method is to partition the yearly LFD (i.e., the "catch curve") using a sequence of cut-off lengths. For a regular series of arbitrarily defined cut-off lengths ($L_c$), corresponding mean lengths ($L_{mean}$) are calculated, i.e., for each $L_c$, the mean length of all fishes that are larger than $L_c$ is calculated. A regression analysis of these transformed data against each other provides an estimate of the intercept (a) and of the slope (b) from a simple linear function of the form $Y = a + b * X$.

The original method, as proposed by Wetherall (1986), plots $L_c$ against $L_{mean}$, and then fits a linear regression on this plot, that always shows a highly significant upward trend (Fig. 1A). Then, it calculates $L_\infty = a / (1 - b)$ and $Z/K = b / (1 - b)$.

In the far more popular "modified" version of this method (Pauly, 1986), a difference ($L_{mean} - L_c$) is plotted against $L_c$ (Fig. 1B). This results in a series of points that always trend linearly downward, which are then fitted with a linear regression, where $L_\infty = a / -b$ and $Z/K = (1 + b) / (- b)$.

*Spurious autocorrelation artifacts – a persistent issue for quantitative science*

Spurious autocorrelation is an artifact that appears in any quantitative study when non-independent variables are used in regression models. Many authors have recognized this as a widespread serious issue for quantitative science, beginning in the late 19th century (Pearson, 1897, Reed, 1921, Chayes, 1949, Bensen, 1965, Kenney, 1982, Kanaroglou, 1996, Brett, 2004, Auerswald et al., 2010). This artifact arises in any regression when the same data are used in the X- and Y-axes, e.g., when performing a linear regression of "(Y = X+E)" against "X". Since the term on the Y-axis contains information from X, this regression will generally be perfectly linear and significant. An in-depth analysis, based on simulations and literature reviews, has been conducted by Brett (2004), who pointed out how widespread spurious autocorrelations are in the modern scientific literature. Despite the long history of these warnings, regression of non-independent variables is still very common in ecology and fisheries science, specifically in length-based studies of natural populations.

*Spurious autocorrelation in the Powell-Wetherall plot methods*

The P-W method is not intended to be used to unveil a relationship between X and Y (e.g., as in Caut et al., 2009, debunked by Auerswald et al., 2010). Still, a relationship between X and Y in the P-W plot is used for all calculations. Therefore, artifacts caused by autocorrelation are not less serious for this method and for any decisions based upon it. As each mean length is calculated, the steadily progressing values of the cutoff lengths are always part of the calculation of mean lengths, so X (cutoff lengths) and Y (mean lengths) are clearly not independent (as required for regression analysis). In the original P-W plot (Fig. 1A), all values used for the Y-axis already contain a substantial amount of information from the X variable (i.e., cutoff lengths are used to calculate mean lengths). That explains why the original P-W plot always points upwards, and regressions are always highly significant.

In the "modified" mP-W version of the plot (Fig. 1B), the strongly autocorrelated construction of the plot is immediately recognizable when looking at the axes titles, where the same parameter appears in both axes. The points used for regression analysis always trend downwards and regressions are always highly significant. Also, the plot is generally being perfectly linear, the data points being even more perfectly aligned than in the "original" version of the P-W plot.

The problem of this artifact is that it produces an apparently perfect structure (a perfect line), independently of the information in the input data. It will always provide the user with the illusion that the model fits perfectly to perfectly shaped data, when actually seeing and analyzing an artifact. In the worst case, the patterns shown in this artifact may be completely independent of data structure. In spite of this potentially devastating (and hitherto ignored) pitfall, this crucial and fundamental aspect of this method has not yet been mentioned or properly tested by any previous studies.

The objective of this study is to assess the effect of spurious autocorrelation on the Powell-Wetherall plot method and to test the reliability of this and other maximum-length-based approaches, by using them to analyze a wide variety of simulated data sets with known input parameters.

## Methods

To test the effects of spurious autocorrelation on the P-W method, synthetic LFD data were used with known input growth and mortality parameters. Estimates for Z/K and $L_\infty$ obtained from P-W and mP-W plots were then compared to known input values. Initially, FISAT II and TropFishR were used to create mP-W plots and obtain these estimates manually one-by-one (with standard errors for $L_\infty$ and K), using manual selection of data points for regression, as widely recommended (Gayanilo et al., 1995, 2005, Sparre and Venema, 1998, Mildenberger, 2017).

To be able to run a large number of Monte Carlo simulations with varying inputs of growth and mortality parameters, an automatic sampling and selection routine was written in R (AR-WethPlot, http://rpubs.com/rschwamborn/267465). In the AR-WethPlot routine, implementations of the "original" P-W and "modified" mP-W plots could be used for comparison. For each run, 10,000 individual growth trajectories (with seasonal recruitment) were generated, sampled monthly and analyzed (Fig. 2).

Several selection strategies were used to test the effect of subjective manual choice of data points for regression (a step that is an intrinsic part of this method). After initial tests, the selection strategy chosen for routine simulations was to select data from L = "mode+10%" to L = "93% $L_{max}$" (= "gamma" selection). This simulates a manual exclusion of the small size classes, that are close to the mode, which may be influenced by incomplete recruitment and mesh selection, and of the largest size classes, which may be difficult to sample quantitatively. This "gamma" selection strategy, similar to what users will do manually in FISAT II and TropFishR, yielded a sample size of at least four data points (size classes) for regression, with usually four to nine perfectly aligned data points used for regression, as done in actual practice. The gamma selection occurs *a posteriori*, after calculation of all mean lengths, just before the regression is performed, as done in FISAT II and TropFishR. Alternatively, an early "beta" selection was also tested, where the data that are not used in the analysis were already excluded from the original LFD data, prior to calculation of mean lengths.

The following types of LFD data were tested:

I.  Initially, simulations were run with ideal, homogeneously sampled LFD data from perfect synthetic populations (as in Fig. 2). Samples were taken from an age-structured, Stochastic Individual-Based Model (SIBM-POP4, http://rpubs.com/rschwamborn/270531), based on well-defined cohorts with inter-annually variable cohort strength and duration of the recruitment season. The duration of the recruitment season in each scenario was calculated as: $D_{rec}$ = 2 * standard deviation of the recruitment curve (in days). All individuals followed a non-seasonal von Bertalanffy growth, with fully stochastic mortality and growth. Several values of K, $L_\infty$ and Z were used. Sampling was performed using a sigmoid gear selection shape, with equally high sampling probabilities for all large individuals (i.e., a "bottom-trawl" selection ogive). For each SIBM-POP4 run, 10.000 individual growth trajectories and mortalities were simulated monthly for ten years.

Samples were only taken when the population was already under full equilibrium, i.e., when less than 1% of the population was derived from the first cohort. This equilibrium criterion was always fulfilled before the onset of the tenth year of simulation (when Z > 0.5). In the last year, samples of the population were taken monthly (e.g., n = 500 ind. sampled randomly each month), before pooling all length samples together, and creating a single annual LFD histogram (i.e., a "catch curve"), as done usually in actual practice. This pooled LFD histogram (e.g., n = 12 * 500 = 6,000 ind. per annual catch curve) was then used for the P-W and mP-W plots.

To assess the effect of random sampling of 500 individuals per month on the spread and bias of estimates, all SIBM-POP4 runs were also analyzed with a mP-W plot of the catch curve of the whole population of the last year (> 20,000 ind.), additionally to the mP-W plot based on the random samples. To compare the results of the SIBM-POP4 runs with previous studies, a more simple, cohort-based synthetic population model (SYNPOP2, http://rpubs.com/rschwamborn/270537) with non-stochastic, deterministic mortality of each cohort (as in Isaac, 1990 and Hufnagl et al., 2013) was also used for comparison.

Yearly cohort strength ($N_{rec}$), K, $L_\infty$, to, and Z were assumed to be random normal variables with predefined, constant variance.

Coefficients of variation (C.V.) of K, Z and $L_\infty$ varied from 5% to 30%, based on the laboratory rearing experiments of Isaac (1990). Monthly sample size varied between simulations, from 200 to 500 ind. month$^{-1}$. Duration of the recruitment season ($D_{rec}$), varied random uniformly from 44 to 263 days between simulations, but was constant within each simulation. Average cohort strength ($N_{rec}$) was usually 500 to 1,000 ind. year$^{-1}$, with an inter-annual C.V. that varied from zero to 50%. $L_\infty$ was set to 10 cm (it does not affect the analyses, since it is merely a scaling factor), K varied from 0.6 to 2, Z from 0.5 to 8, Z/K ratios varied from 0.8 to 7 (Beverton and Holt, 1956; Pauly, 1998, Palomares and Pauly, 2009a, Hufnagl et al., 2013).

Several combinations of input values were tested in three experimental approaches (exp. Ia to Ib). In the first experiment (Ia), all parameters were kept at optimum (i.e., all input parameters were fixed at values which produced minimum bias in output estimates) and up to 500 replicate runs were conducted. Then (Ib), each parameter was varied one by one (e.g., 500 runs with varying Z/K inputs, while all other inputs were fixed at their optimum values). Finally (experiment Ic), a single multivariate analysis was performed where all input parameters were varied at random, using up to 2,000 unique combinations of 12 input parameters.

II. Simulations as in I, but sampled with a bi-sigmoid ogive ("gill-net"-type selection). This simulates common under-sampling of large-sized individuals, which may be due either to gear selection (e.g., more effective gear avoidance by fast-swimming large individuals or due to the use of gill nets) or due to the emigration of large individuals to other areas. Two types of experiments were conducted: IIa: reduction of the amount of large individuals, without creating "zero" values ($L_{max}$ remained unchanged, Fig. 4) and IIb:, complete removal of large size classes ($L_{max}$ was reduced, Fig. 5). Well-defined cohorts, non-seasonal von Bertalanffy growth and varying input values, all settings as in simulations Ia to Ic.

III. Synthetic data taken randomly from LFD data with a linearly declining shape (Z/K = 2), as defined by its intercept (a'), slope (b') and error term (e'), to test the effects of linear and error terms (variability) on the results (Fig 6).

IV. Synthetic data with no growth and zero mortality (i.e., sampling at one occasion only), taken randomly from normally distributed ("bell-shaped") LFD data (Fig 6). Since there cannot be any "true" underlying Z/K ratio or $L_\infty$ for such data, these experiments were conducted solely to test the effect of such common shapes and their standard deviation on the output estimates.

V. Synthetic data with no growth and zero mortality, taken randomly from uniformly distributed ("flat") LFD data (Fig 6), to test the effect of such shapes on the output estimates.

Initially, analyses were performed with a manual selection of "best fitting points" for mP-W plots, using FISAT II, TropFishR and an automatic routine ("AR-Wethplot") with gamma selection, to test the robustness of the AR-Wethplot routine and possible bias due to subjective data point selection, using fifty different datasets among the types of data listed above.

Bias (B, in %) of estimates obtained for $L_\infty$ and Z/K with this method was calculated as:
B = 100 × (estimate - input) / input. Frequency distributions of Bias were compared between scenarios, to find the "ideal" scenario, that gives the most accurate results. For the sake of simplicity, the results (P-W estimates) of each run were classified into four discrete bias categories: 1.) "reasonably accurate" (bias < 10%), 2.) "moderate bias" (30% > bias >10%), 3.) "severe bias" (bias > 30%), 4.) "error" (no estimates could be obtained, e.g., when there are not enough data points available after the mode of the catch curve). For each scenario, the results were summarized into these four categories. Overall, more than 2,000 synthetic populations (see above) were analyzed using the P-W method, based on a total of more than 20,000,000 individual growth trajectories.

## Results

*Relationship between $L_\infty$ and $L_{max}$*

Among the synthetic populations obtained in this study (all with input $L_\infty$ = 10 cm), $L_{max}$ varied whithin a wide range, between less than 7 cm and more than 20 cm (Fig. 2). For large samples and high intra-cohort variability in $L_\infty$ ("Type A" populations, e.g., C.V.$_{L_\infty}$ above 15%), $L_{max}$ was always much higher than the average $L_\infty$ of the population (Fig. 2A). In populations with slow growth and very high input Z/K ratio ("Type B" populations, Fig. 2B), $L_{max}$ was always far below $L_\infty$ (Fig 2B). In summary, the output $L_\infty$ / $L_{max}$ ratio depended on input growth and mortality parameters, sample size and intra-cohort variability in $L_\infty$ (Fig. 2).

*Precision of different P-W methods under optimum settings*

Linear regression models constructed with both plot methods ("original" P-W and "modified" mP-W plots) always produced identical results, whether using FISAT II and TropFishR (with several scenarios of manual choices of points) or the automatic routine (AR-Wethplot) that was used for simulations. The two plot methods and the three applications, when fed the same data points, were identical within reasonable decimals, showing an extremely high precision and correct calculations.

For perfect synthetic data with input Z/K ratio close to 2, estimates of $L_\infty$ and Z/K were extremely robust to changes in the selection of data points (when the "n" of available data points was at least four). Estimates were virtually identical, whether selecting all data above the LFD mode, data above the mode+10%, above the mode+20%, or from mode+10% to 93% $L_{max}$ (identical within 5.6% for $L_\infty$ and within 3.9% for Z/K). This extreme precision and robustness to any data selection strategy is due to the often observed absolutely perfect linearity of all plotted points, above the LFD mode, when using perfect synthetic data with low variability and optimum settings (Fig. 2C).

The "gamma" exclusion of data (e.g., from "mode+10%" to "93% $L_{max}$") *a posteriori* (after calculation and plotting), just before the fit of the regression line (as done in routine analysis) had no relevant effect on the estimates, for most synthetic data. Any exclusion done *a priori* of the smallest size classes had no effect on the results, either. Conversely, any exclusion of large size classes done *a priori* on the original LFD ("beta" exclusion), drastically affected the results (Z/K and $L_\infty$). Since the P-W estimate of $L_\infty$ is strongly influenced by $L_{max}$, the *a priori* exclusion of the largest sizes classes from the data reduces $L_{max}$ and thus proportionally reduces the resulting $L_\infty$ estimate. The Z/K estimate is even more drastically reduced (by up to 50%) than $L_\infty$, when *a priori* excluding the largest size classes from the analysis.

*Accuracy of P-W methods under optimum settings*

The accuracy of P-W methods was tested using synthetic populations with known input parameters. Optimum results (i.e., less biased estimates) were obtained when sampling the whole population (sample = max.) with a recruitment season of 98 to 150 days per year (optimum: 131 days), Z/K of 1.6 to 2.4 (optimum: 2.2), K above 0.6, $C.V._{L\infty}$ below 15%, $C.V._K$ between 5% and 15%, and minimum $C.V._Z$. The optimum Z/K ratio of 2.2 (minimum bias in Z/K estimates) coincides with a very linear shape of the descending part of the catch curve (Fig. 2C), assuming very low intra-cohort variability (C.V. = 5%) in Z, K, and $L_\infty$.

Under these unique optimum settings (simulations Ia), bias in Z/K and $L_\infty$ was generally acceptable (within 10%) in the vast majority of runs (Fig. 3). Any deviation from these ideal settings (e.g., any change in input Z/K values or any increase in intra-cohort variability), leads to severe increase in bias, especially for Z/K (Figs 2, 3 and 4).

Accuracy of Z/K estimates obtained with the P-W method was very sensible to changes in input parameters during experiment Ib, especially to variations in input Z/K ratio, duration of recruitment season, C.V.s of K, $L_\infty$ and Z. For instance, an increase in $C.V._{L\infty}$ above 15% produces severe bias (>30%) for Z/K and $L_\infty$ estimates, even under otherwise optimum conditions (Fig. 3).

*Accuracy of P-W methods under variable inputs*

Simulations with all parameter combinations under optimum and non-optimum settings (Fig. 4) showed that the P-W method does not produce reasonably accurate estimates for Z/K and for $L_\infty$ for the vast majority (85%) of parameter combinations, during experiment Ic. When considering all parameter combinations (945 successful runs), only 15% of runs produced reasonably accurate Z/K estimates (within 10% bias). Only 44% of runs produced results within acceptable bias for Z/K (within 30% bias). Considering all parameter combinations, only 11% of runs were reasonably accurate in $L_\infty$ (within 10% bias), and only 58% of runs were within acceptable bias in $L_\infty$ (within 30% bias). Additionally to the immense spread in bias, a considerable systematic positive bias (overestimation) of $L_\infty$ and Z/K was observed. In this set of simulations, the vast majority of bias was positive (Fig. 4), mainly due to the effects of variability in the original data, which the Wetherall (1986) equations did not account for. Confidence intervals of estimates obtained with the P-W method did often not include the initial input values for $L_\infty$ and Z/K, and were always absurdly narrow for Z/K (Table 1).

Accuracy of $L_\infty$ estimates provided by the P-W method was most strongly determined by the intra-cohort variability in $L_\infty$ (C.V.$_{L_\infty}$). When increasing intra-cohort variability in $L_\infty$ (e.g., increasing input C.V.$_{L_\infty}$ from 5% to 30%), this always increased $L_{max}$, and thus the estimate of $L_\infty$ and the bias in $L_\infty$. This indicates that $L_\infty$ estimates may be systematically biased towards $L_{max}$, even for ideal simulation scenarios. Increasing input C.V.$_{L_\infty}$ clearly increased the bias of $L_\infty$ estimates (Fig. 4). P-W estimates of $L_\infty$ in simulations with input C.V.$_{L_\infty}$ above 7% were not reasonably accurate (bias > 10%) in the majority of runs. C.V.$_{L_\infty}$ above 15% led to severe bias (bias > 30%) in most runs (Fig 4). Furthermore, high values (above 20%) of C.V.s of $L_\infty$ and K resulted in considerable percentages (up to 25%) of simulations with error, i.e., "exotic" catch curves (e.g., extremely skewed shapes, with less than four points for regression), that were not considered in this study.

When using ideal, perfectly sampled synthetic populations, changes in monthly sample size within the tested range (e.g., above 200 ind. per month) had a negligible effect on the accuracy of the P-W methods. This is explainable by the perfectness of random sampling in this study, the large number of individuals sampled per year (above 200 × 12, i.e., above 2,400 ind. per simulation) and the dominance of other stochastic and systematic bias effects that lead to severe bias, prior to the introduction of sampling error.

Clearly, this method is not suitable for assessing Z/K and $L_\infty$ for populations with completely unknown Z/K rates, unknown $L_\infty$ and unknown C.V.$_{L_\infty}$, even when assuming an ideal situation with perfect, non-seasonal VBGF growth, non-seasonal mortality, and perfect representation of large individuals.

*Influence of rare, large-sized individuals on Z/K and $L_\infty$ estimates*

Reducing the amount of large-sized individuals in the sample (experiment IIa) had a significant effect ($p < 0.0001$, Mann-Whitney U test) on $L_\infty$ and Z/K estimates (Fig. 5). This reduction changed the shape of the catch curve, without affecting $L_{max}$. P-W estimates of Z/K ratio were, as expected, dramatically affected by any change in the shape of the catch curve (with bias above 30% in virtually all runs). Reduction in the catch of large size classes leads to a significant ($p < 0.0001$) change in $L_\infty$ estimates given by the P-W method, but the magnitude of change in estimated $L_\infty$ was still within 10% bias, as long as no length classes were completely removed, i.e., as long as $L_{max}$ remained unchanged.

The complete removal of the largest size classes (i.e., reduction of $L_{max}$, experiment IIb) had a severe effect on Z/K and $L_\infty$ estimates (Fig. 5, see "catch prob. = zero"), leading to severe bias ($bias_{L_\infty} > 30\%$). For example, when $L_{max}$ was manually changed from 9.5 cm to 7 cm (Fig. 5), the median $L_\infty$ estimate was reduced from 9.4 cm to only 6.8 cm ($Bias_{L_\infty}$ median: -32%, 95% $Conf.int_{Bias}$: -34.2 to -31.0% ). Thus, the presence/absence of a single large individual in the sample may affect the accuracy of this method regarding $L_\infty$ under such conditions. Simulations with different $L_{max}$ (different absentee size classes, as in Fig, 5) showed very clearly that $L_\infty$ estimates provided by the P-W method are less sensible to the reduction of catch in the large size classes, but are very closely tied to $L_{max}$, for this set of simulations. Thus, when comparing ideal and bi-sinusoidal data, it becomes clear that this method can be very sensible to even minor changes in the occurrence of very large individuals, which are often rare and difficult to capture in a consistent way. The occasional appearance of one single large-sized individual in the sample may thus drastically affect the estimates of Z/K and $L_\infty$ obtained with this method, with no possible correction for this. Exclusion of the largest size classes from the regression ("gamma" exclusion) or weighing by abundance in the ecosystem does not solve this fundamental problem, since the shape of the whole linearized plot is drastically affected by changes in the abundance and occurrence of large individuals.

*Tests with other types of data*

The analysis of perfectly linear "textbook" catch curves (Z/K = 2) with downtrending linear shapes (Fig.6), always led to Z/K estimates that were close to Z/K = 2, as expected. All ideal linear downtrending catch curves were set to converge towards $L_\infty$, i.e., catch was set to be zero at $L_\infty$ ($N_{L_\infty}$ = zero). Analyses of such ideal LFD data with P-W methods yielded, as expected, extremely accurate estimates. Tests with linear catch curves and using different combinations of input slopes and input error terms (random normal variability around straight linear catch curves, experiment III) always produced Z/K estimates of approximately 2 (median: 2.05, mean: 2.04, st. dev.: 0.08, N = 5000 runs), when simulations were run based on catch curves with ideal linear shapes, with negative slopes and $N_{L_\infty}$ = zero. $L_\infty$ estimates were always within 10% of input $L_\infty$ (median: 10.2 cm, mean: 10.24 cm, st. dev.: 0.12 cm, N = 5000 runs) in these ideal linear simulations.

The response of mP-W plots to several types of non-ideal linear shapes (Fig. 6) was tested with 5000 unique combinations of a (intercept, from 3 to 50) b (slope, from -30 to 30) and e (st. dev. of error term, normally distributed, 0 to 20), "x" (i.e., length) values from 0 to 10 cm, and full data selection (no "gamma" selection) of points), which were, in their vast majority, clearly inappropriate to describe Z, K or $L_\infty$ of any natural populations (experiments III and V). Nevertheless, all 5000 runs resulted in nice-looking linear mP-W plots that would give the user an impression of extreme accuracy of this method, even for such clearly inappropriate data. All runs yielded perfectly linear downward mP-W plots, even for completely flat shapes (slope = zero), for fully random data clouds (high input error terms), and even for positive slopes (positive b, low error term). $L_\infty$ estimates were clearly determined by $L_{max}$. The vast majority (>95%) of $L_\infty$ estimates provided by mP-W plots were within 10% of $L_{max}$, independently of the slope of the catch curve (median $L_\infty$ estimates: 9.8 cm, mean: 10.6 cm, st. dev.: 14.7 cm, N = 5000). Extreme $L_\infty$ estimates (> 3 × st. dev.) were all generated by flat slopes (> -10) and high input error terms. Z/K estimates based on these random linear catch curves were very variable (Z/K values up to 439), with most values (>90%) within 0.1 to 1 (median: 0.58, mean: 0.69, st. dev.: 5.9, N = 5000).

When testing perfectly "bell-shaped", normally distributed, LFD data (simulating single-sample catch curves, e.g., samples of one cohort taken during one single month, experiment IV), and "gamma" selection of points, most runs also resulted in perfectly linear downward trending mP-W plots, with very variable Z/K estimates of up to 57 (Fig. 6). Among 500 simulations with different st. dev., 50% of runs produced Z/K estimates between 4.9 and 7.5. Average and median Z/K values were around 6 (median: 6.00, mean: 5.67, st. dev.: 18.6, N = 500 runs). Thus, the analysis of bell-shaped catch curves (being sampling-error dominated, intracohort-variability dominated or mortality-dominated) produced Z/K estimates, which are on the "high end" or above the spectrum reported in the literature. Z/K estimates were not significantly dependent on the standard deviation of the catch curves ($p > 0.9$, linear regression analysis, N = 500 runs). Thus, this method will produce apparently useful output estimates for any such data. Catch curves with shapes that are dominated by normally-distributed variability will generally lead to very high Z/K estimates, independently of their standard deviation. Conversely, $L_\infty$ estimates produced by mP-W plots significantly increased, as expected, with the standard deviation of bell-shaped catch curves (linear regression, $p < 0.001$, N=500). Also, $L_\infty$ estimates significantly increased with $L_{max}$ (linear regression, $p < 0.001$, N=500). $L_\infty$ estimates were always above $L_{max}$ (on average, 25% above $L_{max}$), for such bell-shaped catch curves.

# Discussion

The present study demonstrates that the P-W approach should not be recommended for the assessment of population parameters. Its inherent bias and misguiding plots and calculations are clearly due to its generally wrong underlying assumptions and due to an artifact, the spurious autocorrelation. The $L_{max}$ approach was also shown to be unreliable, since there cannot be expected any particular relationship *a priori* between $L_\infty$ and $L_{max}$ for a given population, without prior knowledge of a population's growth and mortality parameters (Fig. 2).

*Estimating Z/K ratios directly from length data in catch curves – an impossible task*

Numerous processes and parameters determine the shape of a catch curve ( Fig, 1C), such as the well-known effects of Z/K ratio and $L_\infty$ (Powell, 1979, Wetherall, 1986, Pauly 1986). Many other processes have proved to drastically affect catch curves, such as intra-cohort variability in growth between individuals (Isaac, 1990, this study), seasonal variability in growth and mortality (Hufnagl et al., 2013), inter-annual variability in cohort strength (Somerton and Kobayashi,1991, Hufnagl et al., 2010, this study), inter-annual variability in growth and mortality (considered to be zero in all simulations, which is unrealistically optimistic), sampling selectivity (this study) and variability generated by stochastic sampling (this study), among many others. Yet, the P-W approach considers the Z/K ratio to be the only determinant factor, which is clearly a wrong assumption, as shown by this and all other studies on this subject (Isaac, 1990, Somerton and Kobayashi, 1991, Hufnagl et al., 2013, this study).

The simulations of Hufnagl et al. (2013) showed that the inclusion of seasonal variability in growth and mortality (to be expected for most exploited populations, including in the tropics) has a devastating effect on Z/K estimates produced by this method. In the present study, seasonal variation was not explicitly considered for the sake of simplicity and as to test this method under ideal conditions, under its ideal premises and assumptions. Including seasonality into the present simulations would have led to even larger bias in Z/K estimates (Hufnagl et al. 2013).

The Z/K ratio is only one among many factors that give birth to the shape of a catch curve, and thus an interpretation of catch curves uniquely under the aspect of Z/K (as done in all studies that use this method), must be unavoidably erroneous. The origin of this misconception is that the P-W method will give any user the impression that catch curves can somehow be "filtered" towards a perfect regression, which will show its main determinant parameter, the Z/K ratio. Yet, the perfect line, the perfect regression, the extremely narrow confidence intervals given by these methods are not based on the shape of the original catch curve, but solely on a statistical artifact, the spurious autocorrelation.

It is probably impossible to determine the exact percent contributions of a variety of violated assumptions and of the spurious autocorrelation effect to the complex multivariate bias distributions shown in this study. However, it seems likely that three implicit assumptions are the main sources of bias: I.) zero variability in growth and mortality, II.) Z/K = 2 (Figs. 3 and 4) and III.) perfect representation of large individuals (Fig. 5). The spurious autocorrelation effect also clearly contributes to the bias (e.g., by "pulling" the estimate of Z/K towards 2 and the estimate of $L_\infty$ towards $L_{max}$), and it is clearly the reason why this method gives apparently useful results, even for the most inappropriate data (Fig. 6). The spurious autocorrelation effect will produce a nice-looking slope in the P-W plots, even when the data are completely randomic, and thus „simulates" a useful linear shape, even when the data are inappropriate and meaningless.

*Extracting $L_\infty$ from catch curves - the importance of intra-cohort variability*

The P-W method tends to produce $L_\infty$ estimates with a bias distribution that is consistently narrower and less biased for $L_\infty$ than for Z/K. However, in this study, $Bias_{L_\infty}$ was unacceptable (above 30%) in several simulations, especially when variability in growth was very high or when large size classes were missing. $Bias_{L_\infty}$ was clearly dependent on $C.V._{L_\infty}$ and Z/K ratio. Even when all large size classes were perfectly represented, $L_\infty$ estimates were very inaccurate in many simulations, when $C.V._{L_\infty}$ and input Z/K ratios were not ideal.

Intra-population variability in growth (i.e., C.V.s of $L_\infty$ and K) has a strong influence on the results of the P-W methods. Values of $C.V._{L_\infty}$ and $C.V._K$ used in this study were rather conservative, considering that C.V.s of more than 40% have been reported (e.g. Soriano, 1990, Marshall et al., 2009). Already such a C.V. as 20% will have devastating effects on the accuracy of this method, both for $L_\infty$ as for Z/K, as already observed by Isaac (1990). Intra-population variability in growth is an undeniable reality that has been often neglected in length-based models.

*The influence of large individuals*

Underrepresentation of large individuals in catch curves is probably a common phenomenon, since these animals tend to be more effective in gear avoidance during sampling. Emigration (e.g., to deeper waters) and selective fisheries mortality (e.g., by manual capture or harpooning of large individuals) are other processes that may lead to the disappearance of very large individuals at rates that are much higher than predicted by average Z. Another, far more complex phenomenon, may be a decrease in mortality with size and age, which can increase the number of large individuals in the population above model predictions. In a very optimistic scenario, both processes may compensate each other. Small changes in the catch data of large individuals (e.g., just a few large individuals more or less in the sample) will have drastic effects on the whole P-W analysis, making this method very sensitive to random and systematic variation in the catch of large individuals.

In the present study, for ideal scenarios (low C.V.s, Z/K below 4, complete length range, etc.), the systematic bias in $L_\infty$ was found to be relatively small (bias within 10%). Thus, under ideal conditions, the P-W method gives a realistic order of magnitude of $L_\infty$, and may thus be useful, when used with caution and awareness of its potential pitfalls, especially regarding a complete representation of large individuals. The exclusion of large size classes during regression (gamma selection) does not solve this problem.

While in FISAT II and in "ELEFAN in R" (which uses the original P-W plot), only the starting point of the regression line can be defined for the plot (with all data points above being automatically selected), in TropFishR (Mildenberger et al., 2017), there is also the possibility of unselecting the largest size classes, which is an apparent advancement, and may apparently solve the problem of the influence of large size classes, at first sight. However, the newly inserted possibility of unselecting the last few data points during the final regression does not solve the problem of the extremely high influence of few large individuals on these methods, since large individuals exert a huge influence the shape of the whole plot, influencing all data points during the process of calculation of mean lengths, and not only of the last points.

The impression of being able to avoid possible bias introduced by undersampling of large size classes by *a posteriori* "gamma" deselection of large size classes is another dangerous illusion presented to the users of this method.

Weighing the points by sample size (Isaac, 1990), or other indicators of a size class' abundance in the ecosystem does not remove the spurious autocorrelation artifact, that is the basis for the linearization process in this method.

*Comparison with other methods*

The P-W method can be summarized as a procedure that provides extremely precise-looking estimates under any circumstances. There are many other autoregressive methods, that are widely used in fisheries science, such as the Ford-Walford plot, the Gulland and Holt plot, etc. (Ford, 1933, Walford, 1946, Beverton and Holt, 1957, Gulland and Holt 1959, Gulland, 1965, Pauly, 1983, Sparre and Venema, 1998). These methods most likely also suffer from spurious autocorrelation artifacts. Yet, the magnitude of these autocorrelation effects and their consequences for aquatic populations are still to be investigated.

One basic shortcoming of the P-W method is that it assumes that all forms of variability in growth and mortality (interannual, seasonal and intra-cohort) are zero. This is simply not possible, even under ideal conditions (e.g. in experimental aquaria and ponds), due to the existence of genetic and epigenetic variability, habitat heterogeneity, and the stochastic nature of feeding, growth and mortality (e.g. Vincenzi et al., 2014, Uusi-Heikkilä et al., 2015). This variability, that is unaccounted for in the P-W model, is one of the main sources of severe bias, as shown in this study. Many other length-based methods also assume zero variability, such as the Beverton and Holt (1956) and the Ehrhardt and Ault (1992) methods. A recent simulation-based study (Then et al., 2015) showed that both methods are severely biased, especially the Ehrhardt-Ault method (bias from -80% to + 140%), even though their simulations only considered very moderate levels of variability (C.V.$_{size-at-age}$ from zero to 9%).

Even for ideally shaped and perfectly sampled populations, estimates of Z/K were drastically inaccurate and strongly biased under most scenarios, confirming the results of previous studies (Isaac, 1990, Hufnagl et al., 2013). Hufnagl et al. (2013) already perceived that this method is not reliable to estimate Z/K ratios. They concluded that this method "led to Z/K 60% higher or lower than the true value". Instead of the P-W method, they suggested to use another length-based method, the LCCC (Length Converted Catch Curve), when $Z < 4.5$ y$^{-1}$. For total mortalities above 4.5 y$^{-1}$, they could not recommend any length-based methods at all, based on their simulations. The LCCC method is apparently more reliable to assess mortalities than the P-W method, which is probably because the LCCC plot does not suffer from any obvious spurious autocorrelation effects. Still, the LCCC method depends on reliable prior inputs of growth parameters ($L_\infty$ and K) to convert lengths to ages, and is probably limited to situations where $L_{max} < L_\infty$. Also, accuracy and error propagation of the LCCC method, starting from the fit of the growth function, is still not being calculated on a regular basis, so that there is still no way to determine its accuracy.

Mark-recapture or length-at-age data can be included and combined with length-frequency analysis (Morgan 1987). Still, this approach is rarely ever used, given the low cost, extreme simplicity and apparent fail-proof dependability of the P-W method.

There are currently no standard methods available to assess C.V.$_Z$, C.V.$_K$ and C.V.$_{L\infty}$ directly from LFDs or age-at-length data. A possible proxy for C.V.$_{L\infty}$ could be the C.V.$_{length-at-age}$ of the oldest individuals, which can be easily calculated (e.g., based on reading of hard structures), but is rarely reported. A recent compilation (Then al., 2015) showed that such C.V.$_{length-at-age}$ values as 7 to 9% may be already considered to be at the "upper range" for exploited fish stocks. The "true" underlying C.V.$_{L\infty}$ (Isaac, 1990) is very difficult to estimate *a posteriori*. Then et al. (2015) used C.V.$_{length-at-age}$ as a proxy for C.V.$_{L\infty}$. However, the C.V.$_{L\infty}$ of the whole population may be very difficult to estimate, since C.V.$_{length-at-age}$ of the oldest individuals is influenced by many factors, including size-selective mortality (Sogard, 1997). Size-selective natural mortality usually leads to a selective removal of small, low-performance individuals ("bigger is better"), starting from the very first larval stages, mainly due to predation and cannibalism in rigorously size-structured ecosystems (Miller, 1997, Sogard, 1997, Grønkjær et al., 2004, Garrido et al., 2015). Conversely, fisheries mortality usually selectively removes the larger individuals (e.g. Shelton et al., 2015). Both size-selective processes together may lead to a narrowing of length-at-age distributions with increasing age. This explains why C.V.$_{length-at-age}$ is often very high for juveniles and young adults and then usually becomes drastically narrower with increasing age (e.g., Schwamborn and Ferreira, 2002, Kendall et al., 2009, Then at al., 2015).

*Consequences of the spurious autocorrelation for stock assessment and conservation*

Globally, most (> 60%) of marine fisheries stocks lack formal stock assessments (Hilborn and Ovando, 2014). There is growing evidence that classic approaches to fisheries management have failed, and that new, more precautionary approaches are urgently needed (Hilborn and Ovando, 2014, García-Carreras et al., 2016). The bias and pitfalls described in this study may be part of the story of declining fisheries, especially for data-poor, tropical stocks, that may still be unassessed or have been evaluated using inappropriate methods, although many other factors may have contributed to these declines.

The results of the present study are extremely relevant for population and ecosystem management, since many studies still rely on the P-W plot as a central method for the study of hundreds of aquatic populations, without being aware of its inherent bias and pitfalls.

It is very difficult to explain how this striking feature (the spurious autocorrelation) of this widely used method could not have been noticed before, especially since this method is being used by a huge community since almost three decades. This is probably due to a deep belief in its "rigorous theoretical approach" (Brey and Pauly, 1986, Brey et al., 1987). This study demonstrates that, instead, it is based on wrong underlying assumptions (e.g., zero variability), and on misguiding plots and calculations.

The potential of misguidance due to the spurious autocorrelation is particularly relevant for critically threatened or overfished species with slow growth and high total mortality, where the largest fish caught can be expected to be much smaller than $L_\infty$ (Fig. 2B), as likely often occurs in the real, severely overfished world (Pauly et al., 1998, Pauly and Palomares, 2005). For such populations, the estimate of $L_\infty$ obtained by the P-W plots will be close to $L_{max}$ (Fig. 2B), which may lead to a severe underestimation of $L_\infty$ and subsequent overestimation of the growth constant K. Grossly overestimating K for such critical populations may have severe and deleterious consequences for species conservation and sustainability of fisheries.

Fitting growth functions to length data usually involves connecting the "peaks" or "modes" of monthly LFDs by the Petersen method (e.g., using ELEFAN I, Pauly and David, 1981, Pauly, 1986). Often, there are several possible ways to connect the modes, e.g., several subjectively possible growth curves that can "explain" the data. Therefore, the use of the P-W plot to restrain these possibilities during ELEFAN I analysis is a crucial part of this method in many real-life situations, where subjectively connecting the monthly peaks does rarely provide unambiguous results for K and $L_\infty$. The use of the P-W plot to constrain these possibilities, by giving one single, apparently very precise value for $L_\infty$, solves this whole problem "miraculously".

As stated by Palomares and Pauly (2009a), the P-W plot may still be useful as a "heuristic", i.e., as one additional method for *a posteriori* confirmation of in-depth analyses, but it has generally not been used as such, rather as the starting point for estimating growth parameters (as recommended by several authors, e.g., Gayanilo et al., 1995, 2005, Mildenberger et al., 2017) or, even worse, as the only method used to evaluate the status of vulnerability for a given species or population (e.g., Silva et al., 2014, Palomares and Pauly, 2009a). Cleary, more independent studies are needed to verify and, if necessary correct, the length-based approaches that are currently being used.

Automated access to large databases, such as FishBase (Froese and Pauly, 2000, 2017), SeaLifeBase (Palomares and Pauly, 2009b) and FishLife (https://github.com/James-Thorson/FishLife) could be used as additional tools for such methods, e.g., by comparing current $L_\infty$ estimates with reliable previous models and historical $L_{max}$ values, on a routine basis. The new package "rfishbase" (Boettiger, 2017) for R is an important step towards such an automated access. Checking $L_{max}$ in a given dataset with other $L_{max}$ values from FishBase or SeaLifeBase may become a routine in future length-based analyses.

This study intends to push the discussion on length-based methods forward, to encourage a fresh look at old paradigms and the development of new, statistically sound models. We are just at the very beginning of a long journey towards reliable length-based methods.

*First Conclusions and Outlook*

1. The P-W approach is clearly not useful to estimate Z/K ratios, even for ideal, perfectly sampled populations. Estimates of Z/K suffer from severe bias under most scenarios, as shown in this and in previous studies (Isaac, 1990, Hufnagl et al., 2013). Bias in Z/K may be as high as 200%, with no simple correction factors available or to be expected.

2. The original P-W method and the "modified" P-W method both suffer from severe spurious autocorrelation. That is the reason why these methods will always produce perfect lines, and output Z/K estimates with extremely narrow confidence intervals, even when fed the most inappropriate data. Thus, they will always provide the users with a false impression of perfect shape and high accuracy, which is a dreadful pitfall.

3. P-W methods may be useful for an approximate assessment of $L_\infty$ only for ideal, perfectly sampled populations, covering the full length range, and only under very specific conditions, such as of low input Z/K and low C.V.$_{L\infty}$ (e.g., below 15%).

4. Furthermore, the estimate of $L_\infty$ is significantly affected by the stochastic and highly variable sampling of large, rare individuals. Manually excluding large size classes after calculation of mean lengths, prior to regression analysis ("gamma" selection), as done manually in TropFishR, does not solve this problem at all. Any subtle change in large size classes will drastically affect the slope of all points presented to the user for selection. For instance, such common phenomena as undersampling (gear avoidance), emigration (e.g., to deeper waters) and higher fisheries mortality of large individuals will lead to significant subestimation of $L_\infty$, to an overestimation of K during subsequent K-scan analysis and finally to overestimation of maximum sustainable yield and underestimation of a population's vulnerability to overfishing, with deleterious consequences for the affected populations and fisheries.

5. For routinely used stock assessment packages, such as FISAT II, "ELEFAN in R" and TropFishR, appropriate warnings (e.g., "Estimates of Z/K are not reliable, estimates of $L_\infty$ are accurate only under specific conditions") should be included in the P-W routines to alert incautious users of the inherent bias and pitfalls.

6. This study unveiled profound conceptual issues in two methods (P-W plot and $L_{max}$ approach) that are commonly used to fix $L_\infty$ during the fit of a VBGF curve to LFD data. Fixing or constraining $L_\infty$ by doubtful methods, as shown in this study, will obviously lead to doubtful growth models. A well-established method for unconstrained search for optimum growth parameters is ELEFAN I (Pauly and David, 1981, Pauly, 1986) with Response Surface Analysis (RSA, Gayanilo et al., 1995). However, the output of the RSA plot is highly dependent on its input grid parameters and this method cannot consider any seasonality in growth nor can it provide any confidence intervals for its optimum output estimates.

New, more robust and reliable approaches are urgently needed for the length-based study of body growth. One promising path is the use of new ELEFAN-based curve fitting algorithms with unconstrained search for optimum combinations of growth parameters, including seasonality (e.g., ELEFAN_GA and ELEFAN_SA algorithms in TropFishR, Mildenberger et al., 2017). Robust confidence intervals for such optimum estimates can be obtained by subsequent bootstrapping, thus considering the inherent uncertainty regarding the underlying data and processes.

**Acknowledgements**

Many thanks to Daniel Pauly for initiating this work many years ago and for numerous inspiring comments. Thanks to M. L. 'Deng' Palomares for encouraging the formation of a new working group on this subject and for her enthusiasm regarding this new project. Many thanks to Marc Taylor and Tobias Mildenberger for their enthusiasm and competence in forming a working group to develop new bootstrap-based methods. Many thanks to Eduardo T. Paes, Matthias Wolff, Tommaso Giarrizzo, Marc Taylor and Tobias Mildenberger for helpful comments regarding this controversial subject.

**Figure Legends**

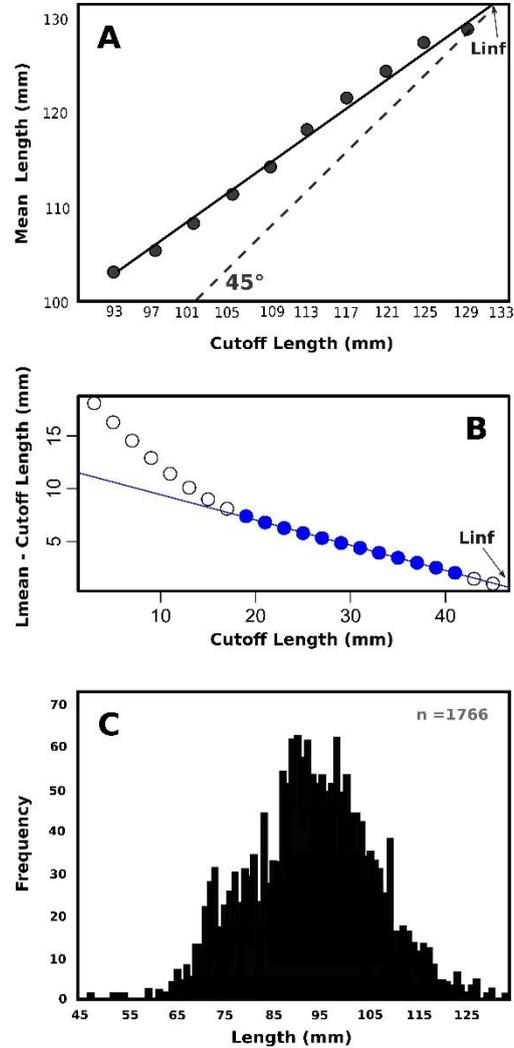

**Fig 1.** The Powell-Wetherall method. A: The original Powell-Wetherall plot (*Wetherall* (1986)) based on the data in "C". B.: The "modified" version of the Powell-Wetherall plot (*Pauly* (1986)), with an example of ideal data, as implemented in FISAT II and TropFishR. C: Annual length-frequency distribution of carapace lengths of rocky lobster (*Panulirus marginatus*) from Maro reef, Hawaii, used as input data for plot "B" (*Wetherall* (1986))

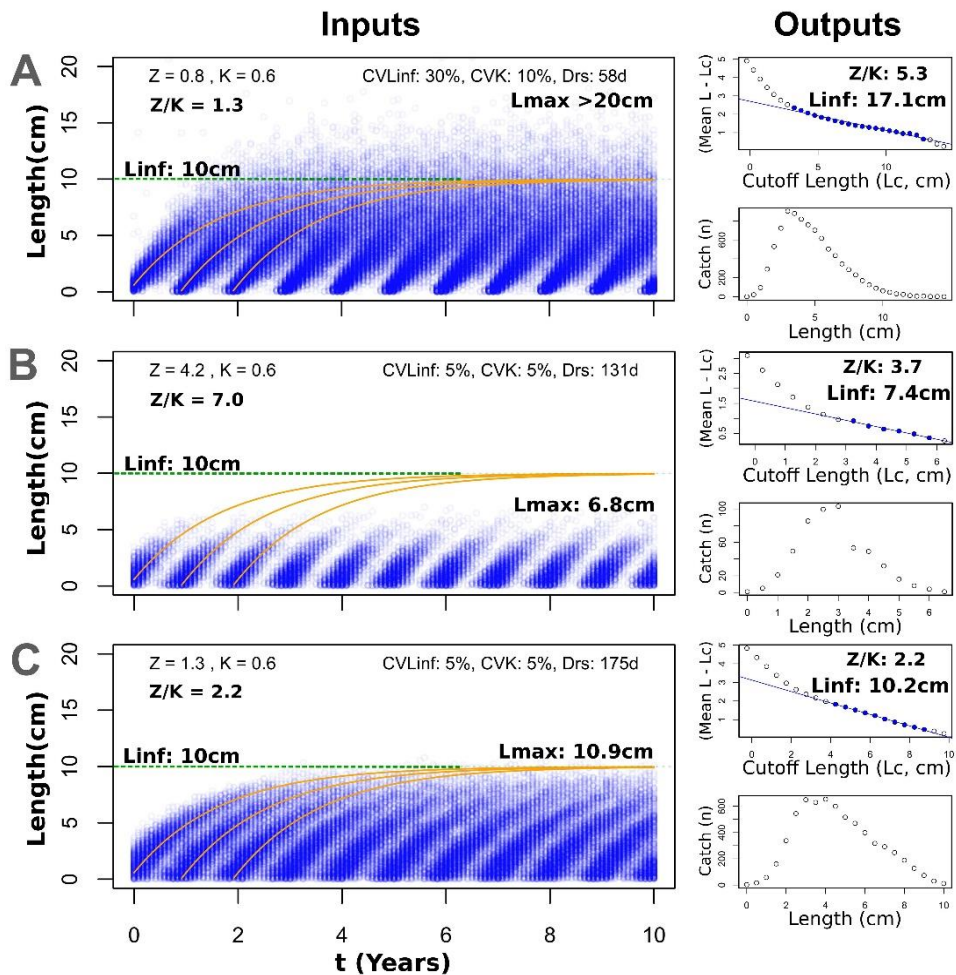

**Fig 2.** Three examples of populations with different relationships between $L_\infty$ and $L_{max}$. Each example is shown with original growth trajectories (left), catch curves and modified Powell-Wetherall plots, with P-W estimates (right). A.) Population with $L_{max} \gg L_\infty$ (high C.V.$_{L\infty}$, low mortality, large sample size, N= 10,000 growth trajectories). B.) Population with $L_{max} \ll L_\infty$ (low C.V. $_{L\infty}$, high mortality, low sample size, N= 5,000 growth trajectories). C.) Population with a combination of parameters that leads to $L_{max} = L_\infty$ (N= 10,000 growth trajectories). All three simulations represent examples of ideal situations, with non-seasonal growth and non-seasonal mortality, zero sampling bias (i.e., these catch curves represent 100% of each population), full equilibrium (zero inter-annual variability in any parameters), and full representation of large individuals in the sampled population (no emigration or gear avoidance by large individuals). Input values used to generate the simulations are given inside the growth trajectory plots (on the left); Powell-Wetherall estimates are given inside the right mP-W plots (on the right).

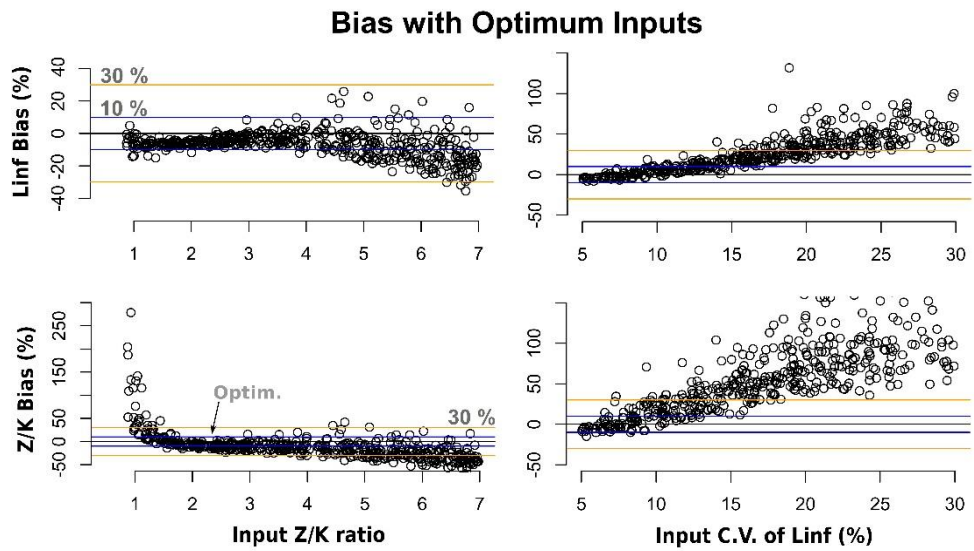

**Fig 3.** Sensibility of the Powell-Wetherall method to variations in input parameters under unique optimum conditions (experiment Ib, see text for details). Each graph shows the influence of variations in input Z/K ratios (left) and in input C.V. of $L_\infty$ (right) of simulated populations on the bias (%) of estimates of Z/K ratios (below) and bias (%) in $L_\infty$ estimates (above) obtained with the modified Powell-Wetherall plot. Orange lines: 30% bias ("severe bias"). Blue lines: limit of 10% bias ("reasonably accurate"). Each graph is based on N = 500 simulations under optimum settings (K = 0.6, C.V.$_K$ = 5%, C.V.$_{L\infty}$ = 5%, C.V.$_Z$ = 5%, non-seasonal growth, non-seasonal mortality, 500 individuals sampled per month, perfect representation of large individuals, recruitment season of 131 days per year, zero inter-annual variability in all parameters, etc.). Gamma selection (excluding large ind.) was used for regression lines in the modified Powell-Wetherall plots. Each point represents the results obtained from a simulation with n = 10,000 individual growth trajectories.

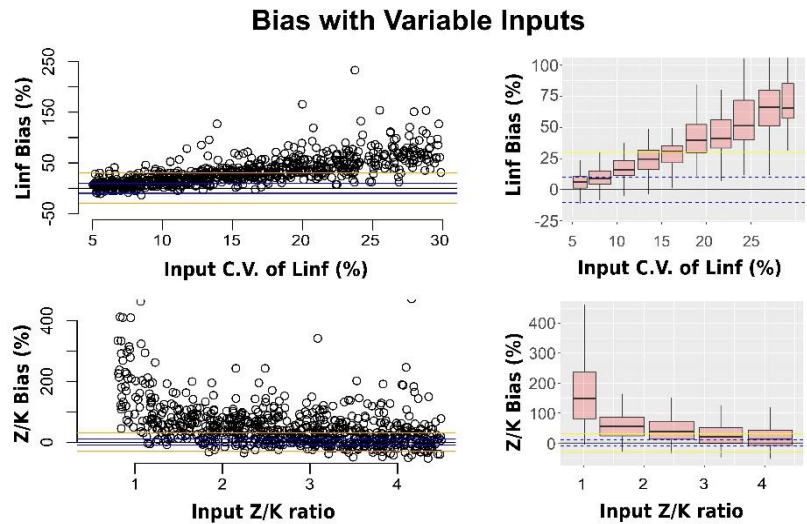

**Fig 4.** Sensibility of the bias in Z/K ratio and $L_\infty$ estimates produced by the modified Powell-Wetherall method under variable input parameter combinations, but under ideal simulations (experiment Ic, ideal, non-seasonal VBGF growth and perfect representation of large individuals, see text for further details). Extreme results (1% most extreme bias values) and "error" runs omitted. Gamma selection (excluding large ind.) was used for regression lines in the modified Powell-Wetherall plots. Orange lines: 30% bias ("severe bias"). Blue lines: limit of 10% bias ("reasonably accurate"). Boxplots: Bias distributions ("zooming" in and summarizing bias distributions). N = 752 successful runs. Each point represents the results obtained from a simulation with n = 10,000 individual growth trajectories.

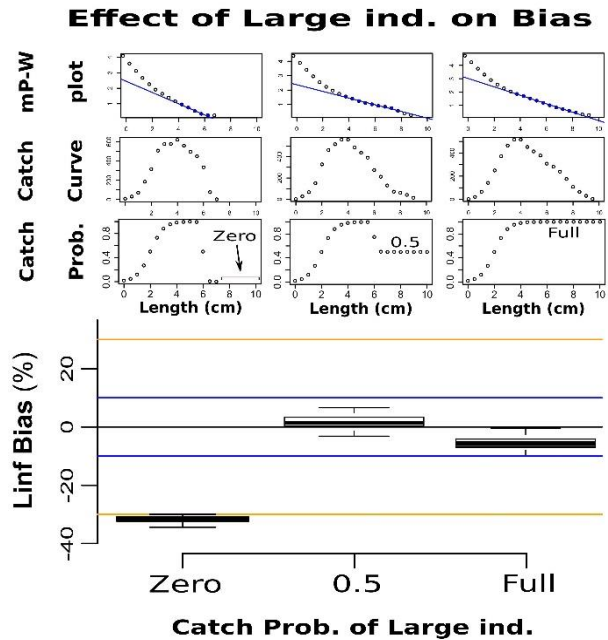

**Fig 5.** Sensibility of $L_\infty$ estimates obtained from the modified Powell-Wetherall method (Wetherall (1986, Pauly (1986) to the reduction of large size classes. All simulations were run with input $L_\infty$ = 10 cm, under optimum parameter settings (Z/K = 2.2 C.V.s = 5%, etc.). Each boxplot represents N = 100 runs with n = 10,000 individuals. Graphs above each boxplot: three selection ogives (catch prob. for each size class), three examples of Catch Curves (examples of catch per size class), and three examples of modified Powell-Wetherall plots (examples of mP-W plots). Full: One set of 100 simulations was run assuming perfect full representation of large individuals in the catch, the other two sets of data show total absence (catch prob. = zero) and partial reduction of catch (catch prob. = 0.5) for large-sized individuals. Gamma selection (excluding large ind.) was used for regression lines in the modified Powell-Wetherall plots. Orange lines in boxplots: 30% bias ("severe bias"). Blue lines: limit of 10% bias ("reasonably accurate").

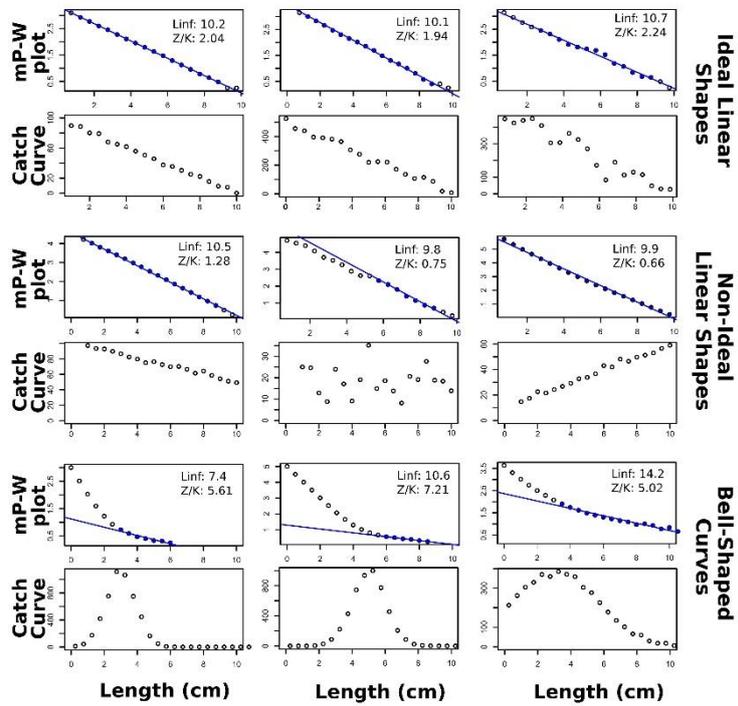

**Fig. 6.** Examples of "modified" Powell-Wetherall analyses (Wetherall (1986, Pauly (1986), based on several types of linear and bell-shaped catch-at-length data. A (above): Examples of ideal linear shapes with downtrending linear shapes. Catch was set to be zero at $L_\infty$ ($N_{L\infty}$ = zero). B (center): Examples of Powell-Wetherall analyses of non-ideal linear shapes. C (below): Examples of bell-shaped, normally distributed data.

33